# Boron Nitride Nanosheets as Improved and Reusable Substrates for Gold Nanoparticles Enabled Surface Enhanced Raman Spectroscopy

Qiran Cai,[1] Lu Hua Li,[1]* Yuanlie Yu,[2] Yun Liu,[3] Shaoming Huang,[4] Ying Chen,[1]* Kenji Watanabe[5] and Takashi Taniguchi[5]

1. Institute for Frontier Materials, Deakin University, Geelong Waurn Ponds Campus, Waurn Ponds, VIC 3216, Australia

2. Advanced Membranes & Porous Materials Center, King Abdullah University of Science & Technology, Thuwal 23955-6900, Kingdom of Saudi Arabia

3. Research School of Chemistry, Australian National University, Canberra, ACT 0200, Australia

4. Nanomaterials and Chemistry Key Laboratory, Wenzhou University, Wenzhou 325027, China

5. National Institute for Materials Science, Namiki 1-1, Tsukuba, Ibaraki 305-0044, Japan

*Corresponding author: luhua.li@deakin.edu.au; ian.chen@deakin.edu.au

Atomically thin boron nitride (BN) nanosheets have been found an excellent substrate for noble metal particles enabled surface enhanced Raman spectroscopy (SERS), thanks to their good adsorption of aromatic molecules, high thermal stability and weak Raman scattering. Faceted gold (Au) nanoparticles have been synthesized on BN nanosheets by a simple but controllable and reproducible sputtering and annealing method. The size and density of the Au particles can be controlled by sputtering time, current and annealing temperature etc. Under the same sputtering and annealing conditions, the Au particles on BN of different thicknesses show various sizes because the surface diffusion coefficients of Au depends on the thickness of BN. Intriguingly, decorated with similar morphology and distribution of Au particles, BN nanosheets exhibit better Raman enhancements than silicon substrate as well as bulk BN crystals. Additionally, BN nanosheets show no noticeable SERS signal and hence cause no interference to the Raman signal of analyte. The Au/BN substrates can be reused by heating in air to remove adsorbed analyte without loss of SERS enhancement.

Introduction

Raman spectroscopy is a valuable non-destructive analytical tool for chemistry, biology, geology, solid-state physics as well as applications in pharmaceutical, cosmetic, food and environment-related industries. However, the very weak signal greatly limits the wide use of this technique.[1] Surface enhanced Raman spectroscopy (SERS) which takes advantage of surface plasmons induced electromagnetic fields and/or chemical charge transfer can enormously increase the signals of Raman spectroscopy and therefore broaden its use,[1-3] especially that the electromagnetic enhancement by metal nanoparticles can achieve single molecule detection.[4-6]

Two-dimensional (2D) nanomaterials, including graphene, hexagonal boron nitride (BN), molybdenum disulfied ($MoS_2$) nanosheets, provide new possibilities for SERS. Actually, graphene has been found to be an excellent substrate for SERS. It can not only boost Raman signals *via* chemical enhancement,[7,8] but also attract analyte molecules in a more controllable way[9] and quench fluorescence.[10] However, chemical enhancement by graphene alone is not as effective as electromagnetic enhancement. Therefore, composites of graphene and metal nanoparticles have also been proposed for SERS to combine the advantages of graphene and the high enhancement factor from metal nanoparticles.[9,11-13] In contrast, the possible use of non-carbon 2D nanomaterials in SERS has not been explored much. For example, there have been only a few reports of using BN nanosheets for Raman enhancement.[8,14,15] Nevertheless, in spite of their similar crystal structure, BN has many different properties from its carbon counterpart, such as distinct chemical bonds and band structure,[16] higher thermal and chemical stability,[17] different surface state and weaker Raman scattering.[18] These properties give BN certain advantages over carbon as a SERS substrate. For example, the stronger resistance to oxidation makes BN nanosheets more preferable for reusable SERS substrate than graphene, because BN nanosheets can sustain heating in air at higher temperatures to remove attached analytes for reuse.[14] In addition, BN nanosheets only show a Raman G band of low intensity,[17-19] which is too weak to show in the SERS spectrum and therefore only Raman signals of analyte are present.

Here, we report that atomically thin BN nanosheets are an excellent substrate for metal nanoparticles enabled SERS. Gold (Au) nanoparticles were prepared on BN by a straightforward but effective and repeatable sputtering and annealing method. The size of the metal particles can be controlled by the thickness of sputtered thin film. Rhodamine 6G (R6G) solution was used to compare the Raman enhancements of Au decorated silicon oxide, atomically thin BN and bulk hBN particles. Interestingly, the atomically thin BN substrate showed the strongest Raman signals of R6G. This phenomenon can be attributed to the unique properties of BN nanosheets. Furthermore, the Au nanoparticles decorated BN substrates for SERS are reusable with no noticeable decrease of Raman enhancement.

**Experimental**

**Preparation of BN nanosheets.**

The BN nanosheets were exfoliated from high-quality hBN single crystals[20,21] on silicon wafer covered by 90 nm thick silicon oxide (SiO$_2$/Si) by the Scotch tape technique.[17,19] Heat treatments at 350 °C were conducted in a tubular furnace in air for 3 h to remove possible moisture and adhesive residue on the samples. The nanosheets were identified by an Olympus BX51 optical microscope equipped with a DP71 camera. Then, a Cypher atomic force microscopy (AFM) was used to measure the nanosheet thickness in tapping and contact modes using Si cantilevers. The nanosheets were also analyzed by a Renishaw inVia Raman microscope equipped with a 514.5 nm laser.

**Synthesis of Au particles.**

Thin layers of Au film (~8-12 nm in thickness) were firstly sputtered on the SiO$_2$/Si substrate with the BN nanosheets and bulk hBN particles under the protection of argon (Ar) at a pressure of ~0.5x10$^{-1}$ mbar (SCD050, Bel-Tec). The sputtering current was 40 mA and the sputtering time was in the range of 20 to 40 s. Then, the Au film covered substrates were annealed at 600 °C for 1 h in Ar atmosphere. The size of the Au particles was measured by the AFM.

**SERS measurements.**

The Au particles covered substrates were immersed in 10$^{-6}$ M R6G (≥95%, Fluka) water (Milli-Q) solution for 1 h, followed by washing with Milli-Q water to remove remained drops of R6G solution. The reusability was tested by heating the substrate with R6G at 400 °C in air for 5 min and re-immersed in the R6G solution of the same concentration for five times. A 100x objective lens with a numerical aperture of 0.90 was used in the Raman measurements. The laser power was ~2.5 mW. All Raman spectra were calibrated with the Raman band of Si at 520.5 cm$^{-1}$.

**Results and discussion**

The BN nanosheets exfoliated on $SiO_2$/Si substrate were heated at 350 °C in air for 3 h to remove possible moisture and tape residue. The moisture may affect the AFM measurements on the thickness of the nanosheets; the tape residue could weaken the adsorption capability of BN nanosheets as well as cause extraneous Raman signals. Figure 1a shows an optical photo of BN nanosheets of 1-3 layers (L) after heating: the thinner the nanosheets, the lower the optical contrast. Consistent to our previous reports,[17,19] the Raman frequency of the G band of atomically thin BN on the $SiO_2$/Si substrate upshifts with the decrease of thickness: 1370.5 $cm^{-1}$ for 1L, 1370.0 $cm^{-1}$ for 2L, 1368.1 $cm^{-1}$ for 3L and 1366.8 $cm^{-1}$ for the bulk. In addition, the width of the G band generally broadens for thinner nanosheets. AFM results in Figure 1c and d show that the thicknesses of the 1-3L BN are 0.40, 0.85 and 1.25 nm, respectively.

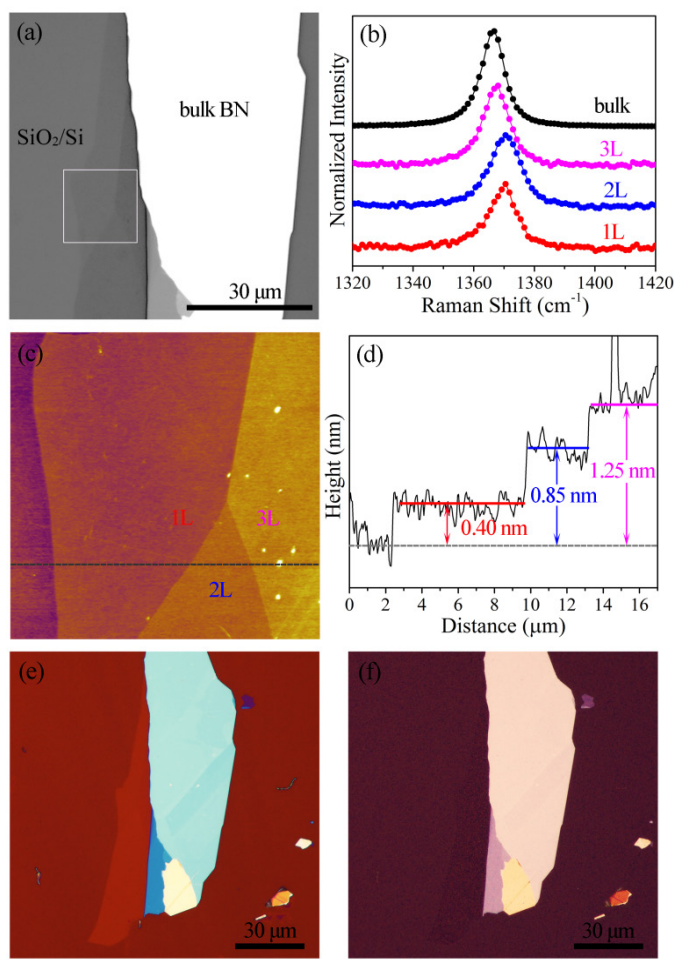

**Figure 1.** (a) Optical image of 1-3L BN nanosheets along with a bulk BN on $SiO_2$/Si substrate; (b) normalized Raman spectra of the BN of different thicknesses; (c) AFM image of the 1-3L BN, as shown in the square in (a); (d) height trace of the dashed line in the AFM image in (c); (e) optical image of the few-layer BN covered by a thin film of sputtered Au; (f) optical image of the same area after annealing to transfer the Au film to particles.

Au nanoparticles were produced on the BN and $SiO_2$/Si substrate by sputtering and annealing method. A thin layer of Au film was deposited on the substrate by sputter coating and then the annealing at 600 °C transferred the film to Au particles. Few-layer BN is visible under optical microscope after sputtering and annealing (Figure 1c and f). It is well-known that surface plasmon resonance and therefore SERS enhancement are highly dependent on the size, shape and distribution of metal nanoparticles. It is found that the size and distribution of Au particles on BN can be easily controlled by this synthesis method. The AFM images in Figure 2a, d and g demonstrate the effect of sputtering time (namely, 20, 30 and 40 s) on the size and distribution of Au particles on BN; while other conditions, such as the sputtering current (40 mA) and annealing temperature (600 °C) are kept the same. According to AFM measurements, the thicknesses of the Au films from the three sputtering time are about 8, 10 and 12 nm, respectively. With the increase of sputtering time, *i.e.* deposition thickness, the Au particles become larger and hence lower in density. The average diameter is 63 nm for the sputtering time of 20 s, 94 nm for 30 s and 154 nm for 40s, respectively (Figure 2b, e and h). The change of particle height follows a similar trend: average 20 nm for the sputtering time of 20 s, 27 nm for 30 s and 35 nm for 40s, respectively (Figure 2c, f and i). Note that the uniformity of both particle diameter and height decreases for longer sputtering time, i.e. broader distributions in Figure 2e, f, h and i than those in Figure 2b and c.

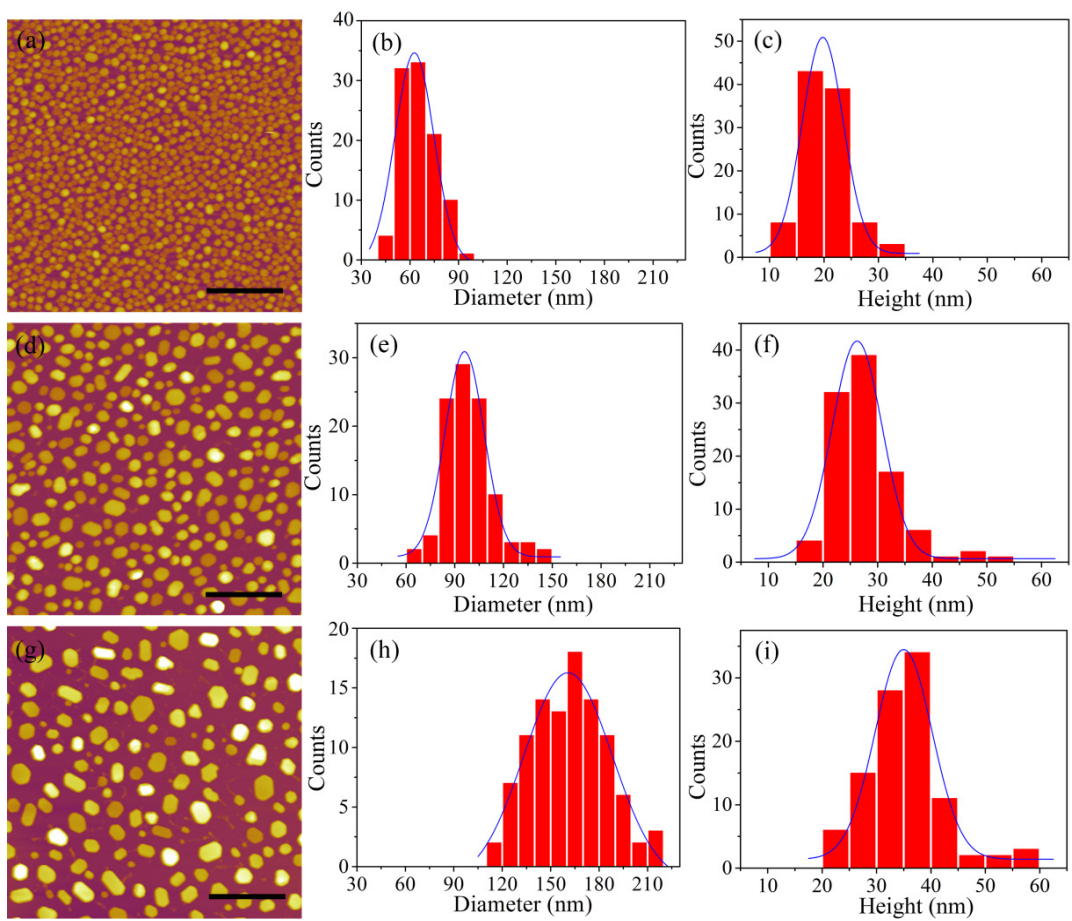

**Figure 2.** (a), (d) and (g) AFM images showing the size and distribution of Au particles on BN using different sputtering time (20, 30 and 40 s, respectively); (b), (e) and (h) statistic of the AFM-derived particle diameter and corresponding Gaussian fits on the distribution; (c), (f) and (i) statistic of particle height and corresponding fits. All scale bars are 500 nm.

Most of the Au particles have a hexagon shape and flat top, suggesting that Au particles are highly crystallized and with a preferential orientation of the (1 1 1) planes parallel to the substrate surface.[13,22,23] This suggests that the Au films melt (at least partially) at the annealing temperature of 600 °C (the melting point of Au could be strongly influenced by its size or thickness)[24] and recrystallized during the cooling process. The different stress and crack in the Au films of different thicknesses during the heating can explain the different particle size and distribution achieved by different sputtering time. The formation of larger particles from the thicker Au film produced by longer sputtering time is due to the tendency of thicker film to break into larger fractions during heating. The less uniformity in particle size and distribution from the longer sputtering time might be attributed to less homogeneity in fractions of thicker film during heating. It should be mentioned that the size and distribution of Au particles can be also adjusted by sputtering current and annealing temperature, which is not shown here.

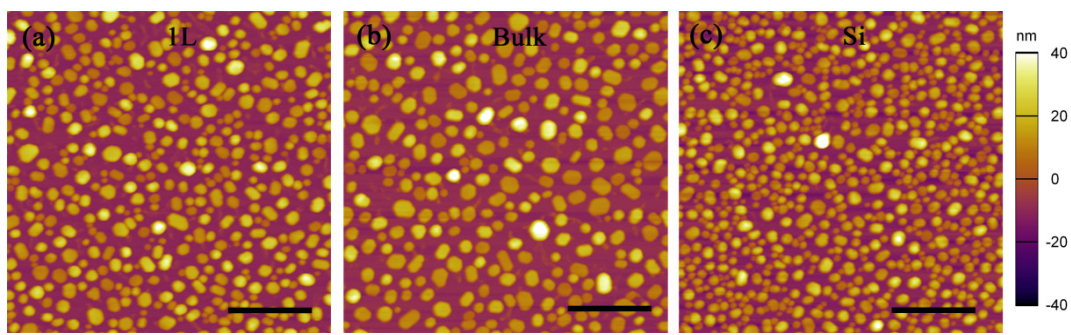

**Figure 3.** AFM images of Au particles produced on (a) monolayer BN, (b) bulk BN and (c) $SiO_2$/Si wafer by 30 s sputtering and 600 °C annealing. The scale bars are 500 nm.

It is also found that under the same sputtering and annealing conditions, the size and distribution of Au particles are slightly different among atomically thin BN nanosheets, bulk BN and $SiO_2$/Si wafer. The Au particles on BN are generally larger than those on the $SiO_2$/Si wafer, but few-layer BN had smaller particles than bulk BN (Figure 3). The particle density on the three substrates follows a contrary trend. The differences are due to a kinetic factor: surface diffusion. As mentioned previously, the Au films melt and hence can diffuse on the substrates during the annealing at 600 °C. Therefore, surface diffusion coefficient, i.e. the rate at which Au migrates and integrates with surrounding clusters to form larger particles, determines the size of the synthesized Au particles. A higher surface diffusion coefficient means a larger migrating rate and hence higher probability to form larger Au particles, and vice versa.[25] It is not surprising that the diffusion coefficient of BN is larger than $SiO_2$/Si,[26] but this is the *first* observation that few-layer and bulk BN have different diffusion coefficients. The difference in diffusion should be caused by different flatness.[27] It is widely known that atomically thin nanosheets are flexible and tend to follow the roughness of the substrate, normally resulting in larger roughness than the bulk crystal.[28,29] The roughness of 1L BN can reduce the diffusion coefficient and result in smaller Au particles.

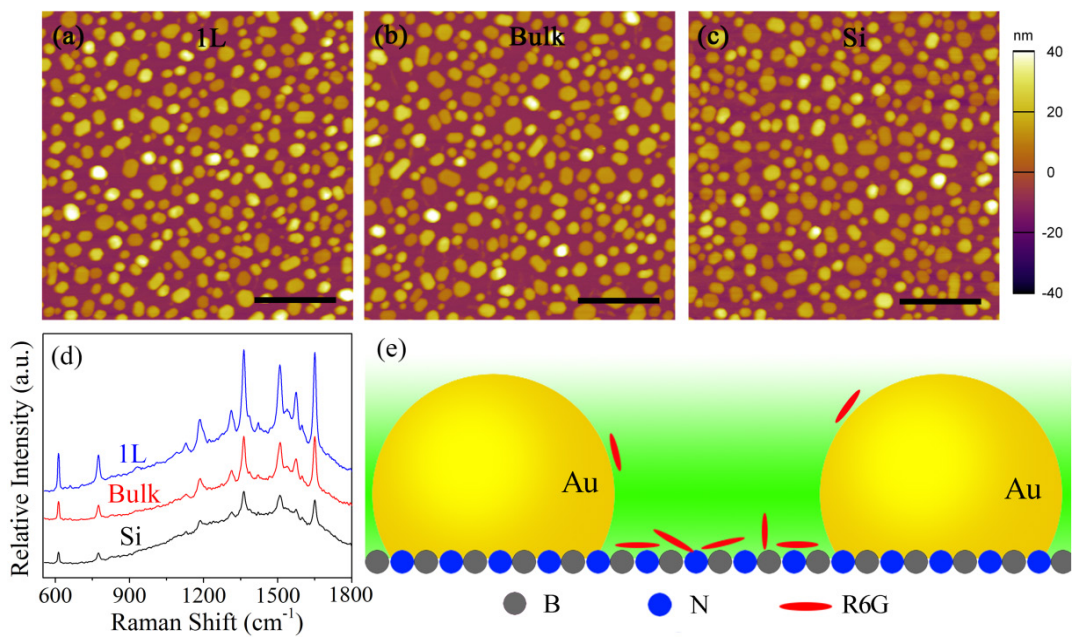

**Figure 4.** AFM images of Au particles on (a) monolayer BN, (b) bulk BN and (c) $SiO_2$/Si substrate, respectively. The scale bars are 500 nm. (d) Raman spectra of R6G on the three SERS substrates; (e) a diagram illustrating that R6G molecules are absorbed on a monolayer BN, which can improve the SERS enhancement.

The size and distribution of Au particles were optimized for SERS purpose. Because the produced Au particles are slightly different for BN nanosheets, bulk BN and SiO$_2$/Si wafer under the same sputtering and annealing condition, sputtering time was tuned to obtain similar size and distribution of Au particles on the three substrates for direct comparison of their SERS enhancements. As shown by the AFM images in Figure 4a-c, similar morphology and density of Au particles were formed on SiO$_2$/Si wafer, 1L and bulk BN using sputtering time of 35, 32 and 30 s, respectively. The statistics on the diameter and height of the Au particles on the three substrates can be found in Supplementary Information (Figure S1). The Raman enhancements of the Au particles decorated three substrates are compared using 10$^{-6}$ M aqueous solution of R6G (Figure 4d). Raman signals of R6G were greatly enhanced by all three substrates, indicating that the faceted Au particles produced by the sputtering and annealing method are suitable for SERS applications. However, both 1L and bulk BN show more signal enhancements than the SiO$_2$/Si substrate, mainly because BN is better in adsorption of R6G via π-π interactions. As illustrated in Figure 4e, although Au particles are able to adsorb a certain amount of R6G molecules, The surface of BN can adsorb dramatically more R6G than SiO$_2$/Si. These additionally adsorbed molecules are located in the so-called "hot spots" or gaps among Au particles and hence greatly electromagnetically enhanced, leading to the stronger Raman signals. The dipole interactions between BN and R6G may also contribute to the better enhancement.[8]

Intriguingly, we noticed that atomically thin BN (including 1-3L) shows more intensified R6G Raman signals than bulk BN particles (Figure 4d and Figure S2 in Supplementary Information). This phenomenon should not be caused by the dipole interaction induced chemical enhancement, because the dipole interactions among BN of different thicknesses should be similar, as demonstrated in a previous study which shows that BN of different thicknesses had the same magnitude of chemical enhancement in Raman.[8] Therefore, this discovery could be due to stronger adsorption capability of atomically thin BN than bulk BN. This special property of BN nanosheets needs further study and may be also applicable to graphene and other 2D nanomaterials.

It should also be emphasized that different from graphene which normally introduces strong intrinsic Raman bands of carbon (G, 2D and possibly D bands) to SERS spectrum,[10,30-32] the Raman signal of BN nanosheets seems barely enhanced by the Au particles and is absent from the SERS results (black spectrum labelled "after heating" in Figure 4d). This makes BN more attractive for SERS, because it does not cause interference with analyte signals.

The reusability of the Au particles coated 1L BN was also tested. To clean off the adsorbed R6G molecules, the substrate was heated at 400 °C in air for 5 min. After the heating treatment, almost no Raman signal of R6G was observed and there was only a featureless fluorescence background (black, Figure 5), implying effective removal of the R6G. The heated substrate was then re-immersed in R6G solution of the same concentration for SERS test, as described previously. The heating and re-adsorption processes were repeated for five times on the same 1L BN decorated by Au particles. Figure 5 shows the Raman spectra from the five cycles of the reusability test. It is worth emphasizing that the 1L BN shows no noticeable loss of SERS enhancement after the five cycles, as there is almost no change in the Raman intensity and the spectral features and peak positions of R6G. The good recyclability can be attributed to the excellent thermal stability of atomically thin BN: monolayer BN can sustain more than 800 °C in air; while graphene starts oxidation at less than 300 °C.[14,17] Thus, the Au/BN hybrid enables reusable SERS substrates that can withstand multiple thermal regeneration cycles.

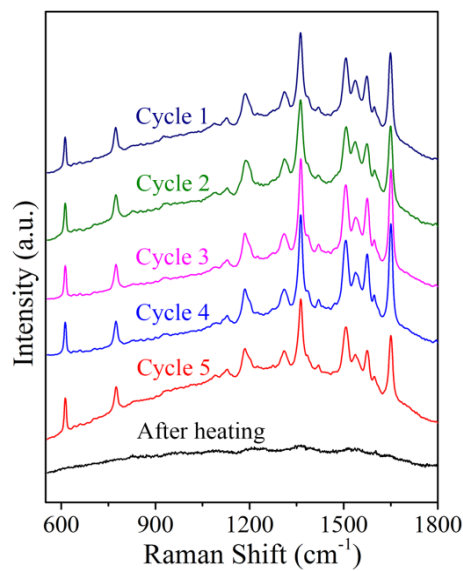

**Figure 5.** Raman spectra of R6G ($10^{-6}$ M) on a SERS substrate of Au decorated 1L BN after five cycles of reusability test. A typical Raman spectrum after heating at 400 °C to remove the adsorbed R6G is also shown for comparison. All the spectra were acquired under the same conditions.

**Conclusions**

In summary, BN nanosheets were used as substrates for highly-sensitive metal nanoparticles enabled SERS substrates. The Au particles were produced by a simple but effective sputtering and annealing method. The particle size and distribution can be controlled by sputtering current, time and annealing temperature, and hence optimised for SERS. It is found that BN nanosheets show better Raman enhancements than bulk BN and $SiO_2$/Si, no interference with Raman signals from analyte and a good reusability without noticeable loss of enhancement. Therefore, BN nanosheets are an excellent substrate for metal particles enabled SERS.

**Acknowledgements**

L.H. Li thanks the financial support from ADPRF2014 and CRGS2015. Y. Chen thanks the funding from the Australian Research Council under the Discovery Program.

# Supplementary Information

**1. Statistics on the diameter and height of the Au particles in Figure 4a-c**

The statistics on the diameter and height of the Au particles on 1L BN, bulk BN and SiO$_2$/Si in Figure 4a-c are shown in Figure S1. It can be seen that both the distributions and averages of the diameter and height of the Au particles on the three surfaces are similar and the direct comparison of their SERS signals in Figure 4d is justified.

**2. SERS spectra of R6G on 1-3L BN**

Figure S2 compares the Raman signals of R6G ($10^{-6}$ M) on 1L, 2L and 3L BN nanosheets covered by Au particles of similar diameters and heights, demonstrating that 1-3L BN nanosheets have a comparable enhancement in Raman signal due to their similar adsorption capability. These enhancements are much stronger than those of bulk BN and the SiO$_2$/Si substrate (Figure 4d).

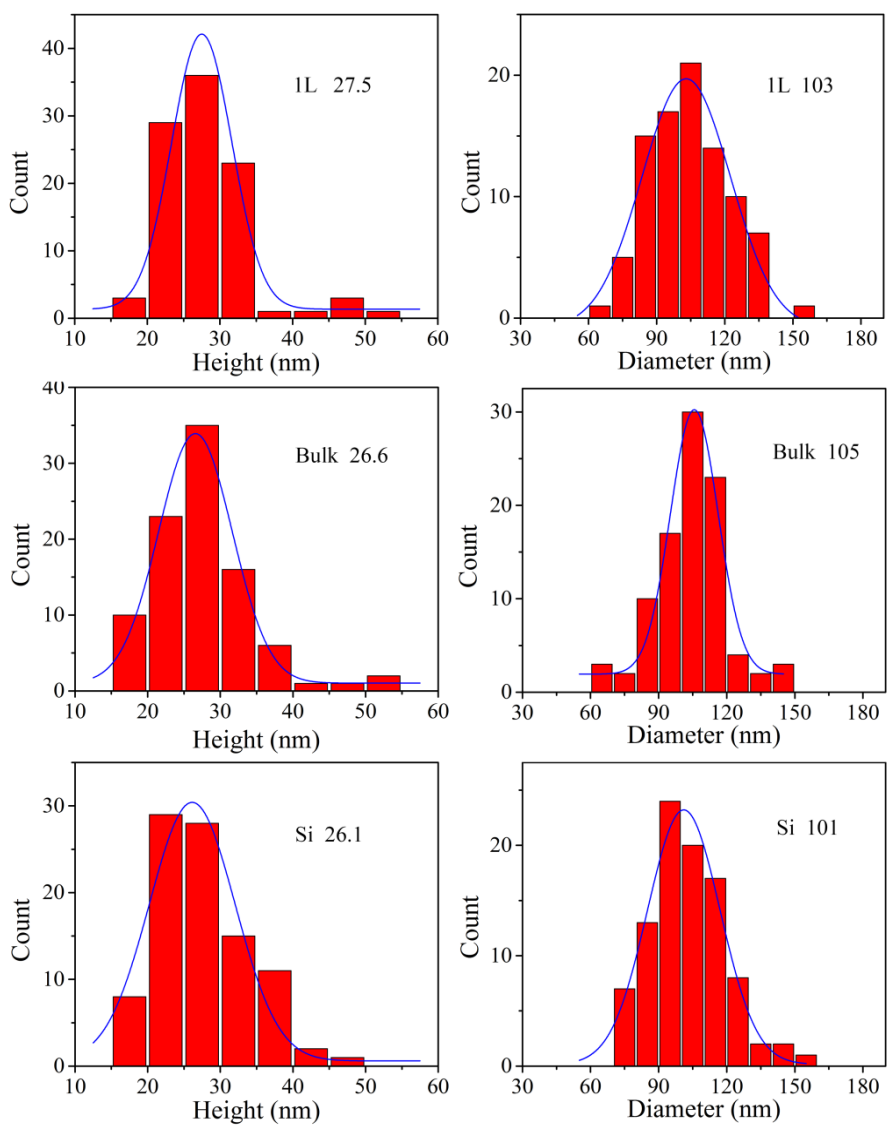

**Figure S1.** Statistics on the diameter and height of the Au particles on 1L BN, bulk BN and SiO$_2$/Si shown in Figure 4a-c.

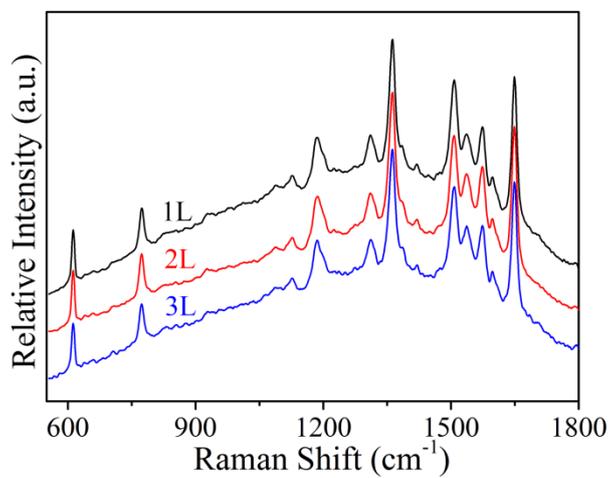

**Figure S2.** Raman spectra of R6G ($10^{-6}$ M) adsorbed on 1-3L BN covered by Au particles of similar diameter and height.